\documentclass[sigconf]{acmart}

\AtBeginDocument{%
  }

\copyrightyear{2024}
\acmYear{2024}
\setcopyright{rightsretained}
\acmConference[CCS '24]{Proceedings of the 2024 ACM SIGSAC Conference on Computer and Communications Security}{October 14--18, 2024}{Salt Lake City, UT, USA}
\acmBooktitle{Proceedings of the 2024 ACM SIGSAC Conference on Computer and Communications Security (CCS '24), October 14--18, 2024, Salt Lake City, UT, USA}
\acmDOI{10.1145/3658644.3691367}
\acmISBN{979-8-4007-0636-3/24/10}
\usepackage{subcaption}
\usepackage{soul}

\usepackage{balance}
\begin{document}

\title{Demo: \texttt{SGCode:} A Flexible Prompt-Optimizing System \\ for Secure Generation of Code}


\author{Khiem Ton*}
\email{kt477@njit.edu}
\author{Nhi Nguyen*} 
\email{yennhi1908hcm@gmail.com}
\affiliation{
\institution{New Jersey Institute of Technology}
\city{Newark}
\state{New Jersey}
\country{USA}
}


\author{Mahmoud Nazzal}
\email{mn69@njit.edu}
\author{Abdallah Khreishah}
\email{abdallah@njit.edu}
\affiliation{
\institution{New Jersey Institute of Technology}
\city{Newark}
\state{New Jersey}
\country{USA}
}

\author{Cristian Borcea}
\email{borcea@njit.edu}
\author{NhatHai Phan} \authornote{Corresponding author; Khiem Ton and Nhi Nguyen are co-first authors.}
\email{phan@njit.edu}
\affiliation{
\institution{New Jersey Institute of Technology}
\city{Newark}
\state{New Jersey}
\country{USA}
}



\author{Ruoming Jin}
\affiliation{%
  \institution{Kent State University}
  \city{Kent}
  \state{Ohio}
  \country{USA}
  }
\email{rjin1@kent.edu}

\author{Issa Khalil}
\affiliation{%
  \institution{Qatar Computing Research Institute}
  \city{Doha}
  \country{Qatar}
}
\email{ikhalil@hbku.edu.qa}

\author{Yelong Shen}
\affiliation{%
  \institution{Microsoft Azure AI}
  \city{Redmond}
  \state{Washington}
  \country{USA}
}
\email{yelong.shen@microsoft.com}


\renewcommand{\shortauthors}{Khiem Ton et al.}
\begin{abstract}
This paper introduces \textbf{\texttt{SGCode}}, a flexible prompt-optimizing system to generate secure code with large language models (LLMs). \texttt{SGCode} integrates recent prompt-optimization approaches with LLMs in a unified system accessible through front-end and back-end APIs, enabling users to 1) generate secure code, which is free of vulnerabilities, 2) review and share security analysis, and 3) easily switch from one prompt optimization approach to another, while providing insights on model and system performance. We populated \texttt{SGCode} on an AWS server with PromSec, an approach that optimizes prompts by combining an LLM and security tools with a lightweight generative adversarial graph neural network to detect and fix security vulnerabilities in the generated code. Extensive experiments show that \texttt{SGCode} is practical as a public tool to gain insights into the trade-offs between model utility, secure code generation, and system cost. \texttt{SGCode} has only a marginal cost compared with prompting LLMs. \texttt{SGCode} is available at: {\color{blue}\href{https://sgcode.codes/}{\textbf{SGCode}}}.
\end{abstract}




\maketitle

\section{Introduction}

Ground-breaking developments of LLM-powered code generation commercial tools, such as Microsoft GitHub Copilot and Amazon CodeWhisperer, have significantly improved software developer productivity \citep{copilot}. 
However; recent studies show that these tools frequently inherit security flaws from their open-source training data, leading to vulnerabilities such as common weakness enumerations (CWEs) in the generated code 
\citep{pearce2022asleep,charalambous2023new,siddiq2023lightweight}. In real-world systems, attackers can exploit such code vulnerabilities, potentially leading to cyberattacks, data breaches, and degraded performance. Therefore, there is an urgent need for reliable approaches and systems to identify and fix vulnerabilities in LLMs-powered code generation.

Recent approaches, including PromSec \citep{PromSec} and SafeCoder \citep{he2024instructiontuningsecurecode}, have shown great potential in optimizing prompts toward generating secure, free of vulnerabilities code. Yet, the lack of a flexible, deployable system limits our understanding of the trade-offs between secure prompt optimization, code utility, security analysis, and system performance.
To bridge this gap, we propose \textbf{\texttt{SGCode}},  a flexible system for deploying and evaluating various prompt optimization methods. This tool enables users to gain insights into the trade-offs between code utility, security analysis, and system performance, ultimately aiding in the development of more secure approaches while minimizing system costs. For example, analyzing different CWEs and their associated costs in computation, communication, and prompting LLMs allows us to refine our model to better address these vulnerabilities and reduce overall costs.

\textbf{Contributions.} We develop \texttt{SGCode} with two integrated components: \textbf{(1) Back-end Services}, which seamlessly integrate \textbf{Code Security Analysis Tools}, such as Bandit \citep{bandit_ref} and CodeQL \citep{GitHubInc2021CodeQL}, and commercial LLMs, such as GPT-4o, to generate optimal prompts for secure code generation and send the results, including the code, security analysis, and system performance report, to the user front-end;
\textbf{(2) User front-end}, a web-based interface that enables users to query prompt-optimizing systems, display code, view and share security analysis and performance reports. \texttt{SGCode} is deployed on an AWS lightweight server, and extensive experiments show minimal system cost (CPU, memory, and latency) compared to the high cost of LLM code generation.
In addition, \texttt{SGCode} allows users to easily switch prompt-optimizing approaches \citep{PromSec,he2024instructiontuningsecurecode}.
\texttt{SGCode} offers a portable and lightweight solution to gain insights into the trade-offs between code utility, secure code generation, and system performance cost. 
\vspace{-5pt}

\section{Prompt-Optimizing Secure Code Generation}


Two recently developed secure code generation approaches are PromSec \citep{PromSec} and SafeCoder \citep{he2024instructiontuningsecurecode}.
The key idea of PromSec is leveraging a graph generative adversarial neural network (gGAN) to mitigate code vulnerabilities by altering the code's graph representations. These altered representations are translated into prompt adjustments using LLMs. PromSec trains the gGAN model using a differentiable contrastive objective, which seamlessly integrates code security analysis tools, LLMs, and the gGAN's generator together, ensuring secure code generation.

An alternative solution is SafeCoder \citep{he2024instructiontuningsecurecode}, an instruction tuning approach to resolve code vulnerabilities in two steps: (1) Curating a dataset consisting of standard instructions and a code vulnerability dataset consisting of GitHub commits for fixing code vulnerabilities; and (2) Instruction tuning that minimizes the language modeling loss $\mathcal{L}^{\text{std}}$ if the data point is in the instruction dataset; otherwise, the loss is $\mathcal{L}^{\text{sec}} + \mathcal{L}^{\text{vul}}$ if the data point is in the vulnerability dataset. $\mathcal{L}^{\text{sec}}$ is the likelihood loss that aligns the LLM model toward the secured tokens in $\boldsymbol{o}^{\text{sec}}$, whereas $\mathcal{L}^{\text{vul}}$ is the unlikelihood loss that helps to reduce the probability of generating vulnerable tokens in $\boldsymbol{o}^{\text{vul}}$. The fine-tuned LLM is used to generate secure code.

\section{\texttt{SGCode} System}

\begin{figure*}[t]
\captionsetup[subfigure]{justification=centering}
\centering
  \begin{subfigure}[t]{0.33\textwidth}
        \centering
        \includegraphics[width=\textwidth, height=0.65\textwidth]{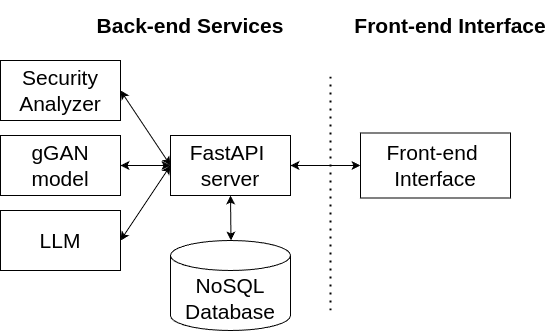}
        \caption{Basic Architecture of \texttt{SGCode}.} 
        \label{fig:architecture}
    \end{subfigure}  
    \hfill
    \hspace{-15pt}
    \begin{subfigure}[t]{0.33\textwidth}
        \centering
        \includegraphics[width=\textwidth, height=0.65\textwidth]{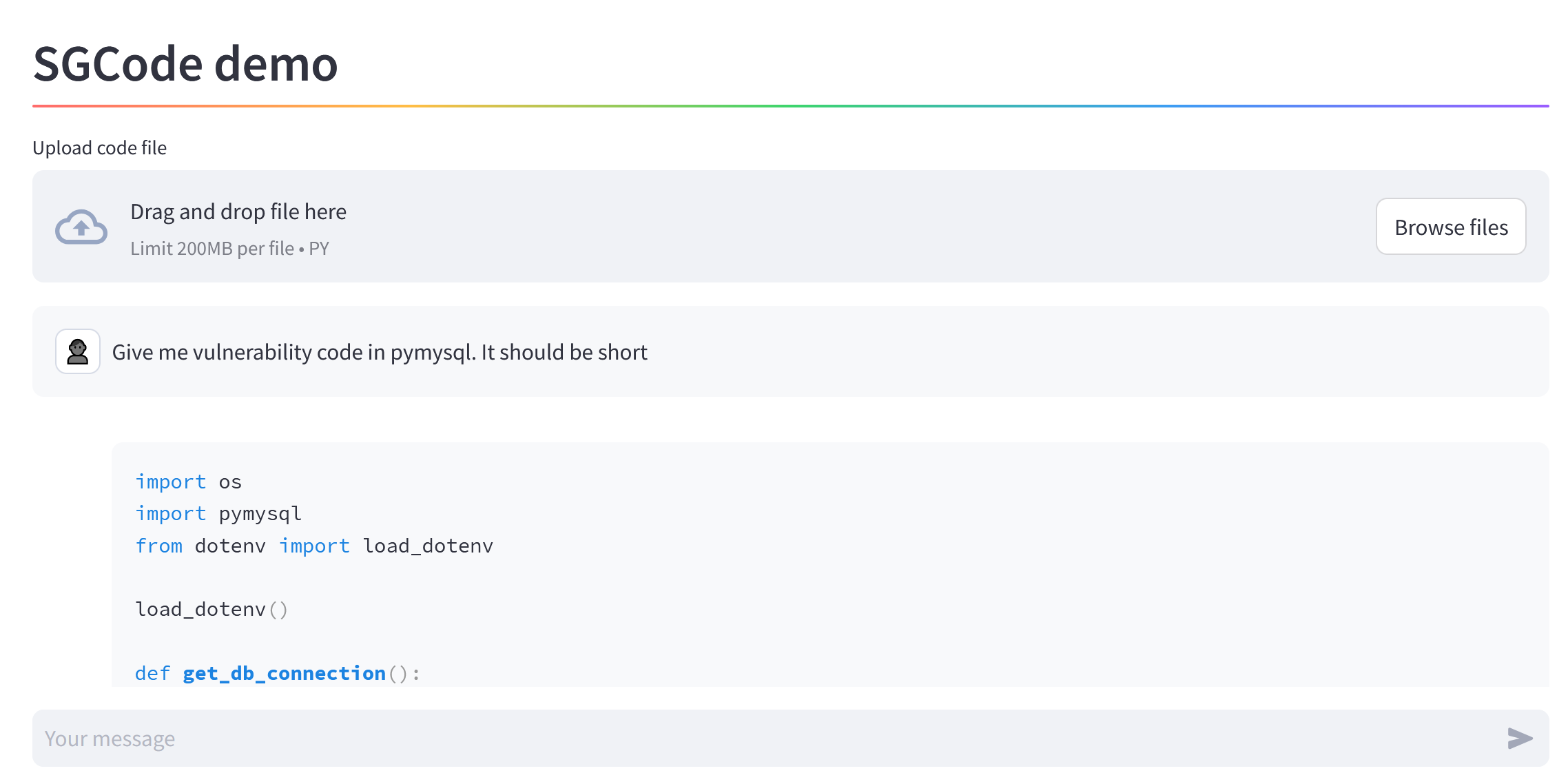}
        \caption{Chat Interface} 
        \label{fig:chat}
    \end{subfigure}
    \hspace{-15pt}
    \hfill
    \begin{subfigure}[t]{0.33\textwidth}
        \centering
        \includegraphics[width=\textwidth, height=0.65\textwidth]{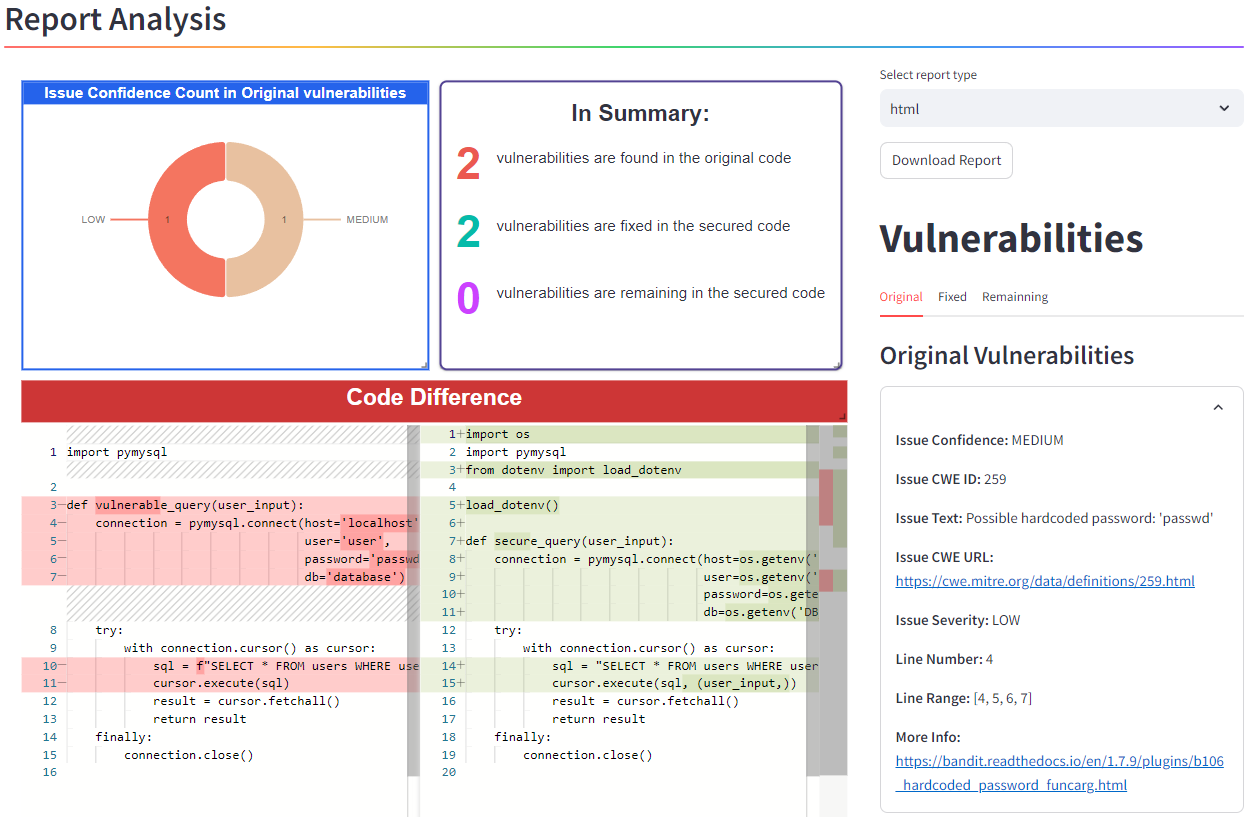}
        \caption{Report Analysis Interface}
        \label{Fig:report}
    \end{subfigure}
    \caption{Front-end User Interface.} 
    \label{Fig:FE}  
\end{figure*}

We develop \texttt{SGCode} (Figure \ref{fig:architecture}) as an API-accessible service and a web application for users to access our system on the web or integrate our API into their applications. 
The system design focuses on simplicity to optimize system costs,  
extensibility to customize the API, and maintainability with our modular architecture. 


\texttt{SGCode}'s back-end is powered by FastAPI \citep{Ramirez_FastAPI}. 
It receives instructions to generate code 
from the users and then utilizes a pipeline consisting of three main components:
(1) an interchangeable security analysis component, (2) a gGAN model from PromSec, and (3) an easily modifiable LLM. This design allows users to generate secure code with customization in mechanisms such as PromSec or SafeCoder. Users can use gGAN with an LLM for PromSec, or they can simply change the LLM to the SafeCoder-tuned LLM as in \citep{he2024instructiontuningsecurecode} to enable SafeCoder standalone or even combine the two approaches. 
Our system employs a NoSQL database to store the generated code for shareable security reports.


Before performing security analysis, the back-end processes the request from the front-end. If the payload instructs code generation, it is fed to the LLM component to produce the code. The generated code is then analyzed for vulnerabilities using Bandit or CodeQL. 

If no vulnerability is found in the code or if the SafeCoder standalone is selected, the system returns the code to the user since the code is secure. Otherwise, the vulnerable code is sent to the gGAN module from PromSec for further processing and then passed to the LLM component. 

The LLM plays a pivotal role in generating code that aligns with users' instructions and performs prompt optimization in PromSec. Our design allows users to modify the LLM by supporting any API servers that are OpenAI API-compatible. As a result, users can use our OpenAI model connections, connect via their personal OpenAI account, or implement self-hosted LLMs. This approach also helps facilitate the integration of SafeCoder into our system.

A FastAPI server manages all three modules, ensuring seamless operation of the pipeline. In addition to managing the modules, the server performs auxiliary tasks, such as saving the programs with their corresponding security analysis to a NoSQL (MongoDB) database and creating shareable reports. These tasks are background processes, thus do not impact the pipeline's performance.

\begin{table*}[t]
    \centering
    \tiny 
    \begin{tabular}{|c|c|c|c|c|c|c|c|c|}
        \hline
        \textbf{\#Vulnerabilities} & 
        \textbf{gGAN time} & 
        \textbf{Security Analysis time} & 
        \textbf{LLM time} & 
        \textbf{Communication time} & 
        \textbf{Total time} & 
        \textbf{\#Prompt Tokens} & 
        \textbf{\#Output Tokens} & 
        \textbf{Time per Code line} \\
        \hline
        2 & 0.01 & 0.97 & 40.28 & 0.62 & 41.89 & 426 & 355 & 0.66 \\
        \hline
        3 & 0.007 & 0.78 & 23.17 & 0.59 & 24.56 & 310 & 264 & 0.53 \\
        \hline
        4 & 0.004 & 0.52 & 11.19 & 0.56 & 12.28 & 318 & 217 & 0.34 \\
        \hline
    \end{tabular}
    \caption{Performance metrics for different numbers of vulnerabilities (time is in seconds).} \vspace{-20pt}
    \label{tab:performance-metrics}
\end{table*}

\begin{table*}[t]
    \centering
    \tiny 
    \begin{tabular}{|c|c|c|c|c|c|c|c|c|}
        \hline
        \textbf{CWE ID} & 
        \textbf{gGAN time} & 
        \textbf{Security Analysis time} & 
        \textbf{LLM time} & 
        \textbf{Communication time} & 
        \textbf{Total time} & 
        \textbf{\#Prompt Tokens} & 
        \textbf{\#Output Tokens} & 
        \textbf{Time per Code line} \\
        \hline
        20 & 0.015 & 1.46 & 55.14 & 0.60 & 57.21 & 365 & 379 & 1.04 \\
        \hline
        78 & 0.007 & 0.77 & 22.30 & 0.58 & 23.65 & 310 & 256 & 0.52 \\
        \hline
        89 & 0.008 & 0.78 & 25.06 & 0.59 & 26.43 & 357 & 294 & 0.52 \\
        \hline
        259 & 0.008 & 0.79 & 25.52 & 0.59 & 26.92 & 339 & 279 & 0.54 \\
        \hline
        327 & 0.005 & 0.58 & 13.26 & 0.58 & 14.43 & 313 & 228 & 0.37 \\
        \hline
        703 & 0.002 & 0.40 & 6.59 & 0.59 & 7.58 & 236 & 207 & 0.22 \\
        \hline
    \end{tabular}
    \caption{Performance metrics for different CWEs (time is in seconds).} \vspace{-20pt}
    \label{tab:performance-metrics-vuln-id}
\end{table*}

\begin{table*}[t]
    \centering
    \tiny 
    \begin{tabular}{|l|c|c|c|c|c|c|c|c|}
        \hline
        \textbf{Prompt Length} & 
        \textbf{gGAN time} & 
        \textbf{Security Analysis time} & 
        \textbf{LLM time} & 
        \textbf{Communication time} & 
        \textbf{Total time} & 
        \textbf{\#Prompt Tokens} & 
        \textbf{\#Output Tokens} & 
        \textbf{Time per Code line} \\
        \hline
        LOW & 0.006 & 0.63 & 15.58 & 0.57 & 16.79 & 282 & 231 & 0.41 \\
        \hline
        HIGH & 0.013 & 1.23 & 51.25 & 0.64 & 53.14 & 471 & 400 & 0.88 \\
        \hline
    \end{tabular}
    \caption{Comparison of performance metrics for LOW and HIGH prompt length (time is in seconds).} \vspace{-20pt}
    \label{tab:performance-metrics-low-high}
    
\end{table*}
\vspace{-5pt}
\subsection{Front-End Interface}

The interface is built using Streamlit, an open-source Python framework for creating interactive user interfaces (UI). The interface is organized into two main screens: (1) AI Prompt UI, which facilitates user input (Figure \ref{fig:chat}); and (2) Security Analysis Report screen, which displays detailed results and insights (Figure \ref{Fig:report}).

\textbf{AI Prompt UI.} Users configure preferences for security analysis tools, code securing mechanisms, and LLMs. Once preferences are set, users can initiate their work by entering a prompt into the input box at the bottom of the screen. In addition, users can upload code directly using the code upload button. The submitted data is then sent to the back-end for processing. For every prompt
for which the back-end service returns a response, the generated code from the LLM is displayed, and a button ``Security Analysis'' is provided to generate the code’s security analysis report.

\textbf{Security Analysis Report.} The report page presents vulnerabilities in the code and provides data visualization. The right side of the screen features a container divided into tabs that distinguish between vulnerabilities in the original code (user input prompt) and those in the secured code (generated by \texttt{SGCode}). Each tab displays a detailed list of issue information. This comparison facilitates the identification of modifications made to address vulnerabilities. On the left side, the page includes a comprehensive visualization of issue confidence counts in the original code, which reflects the severity of vulnerabilities. A summary tab provides an overview of detected vulnerabilities, detailing the number of identified, fixed, and remaining issues, thus offering a rapid assessment of the code's security status. The page also features a line-by-line comparison of the original and secured code to illustrate the changes implemented and the effectiveness of the system’s fixes. Users can share the security analysis as a PDF report or HTML link.

\vspace{-10pt}

\subsection{Deployment and Evaluation}

We deploy \texttt{SGCode} on an AWS c7g.large virtual machine consisting of 2 vCPUs of AWS Graviton3 ARM processor and 4 GiB memory. We run the FastAPI with 4 workers and Streamlit on the same machine. Our back-end connects to a MongoDB instance hosted via MongoDB Atlas.
We conduct three experiments using the test data in \citep{PromSec}: \textbf{(1)} Evaluating \texttt{SGCode}'s resource usage with and without PromSec; \textbf{(2)} Measuring \texttt{SGCode}'s latency given the number of CWEs, CWE IDs, and the prompt length; and \textbf{(3)} Inspecting the code security and code functionality using our Security Report.

First, we measure the resource usage by using the Python libraries psutil and sys. We select the back-end and front-end process IDs and their child processes and run the test dataset on the front end via Selenium. We evaluate our system until the test is finished and report the average result in three runs. Table \ref{tab:resource-usage} shows that utilizing PromSec has negligible CPU and memory usage.


Second, we measure the system's total latency and the latency of each component. We conduct the latency experiment on the vulnerabilities count, different CWEs, and prompt lengths based on the average number of tokens in the test dataset. In our dataset, we consider ``LOW'' to be less than and ``HIGH'' to be larger than the average number of tokens (i.e., 335.5). 
Tables \ref{tab:performance-metrics}-\ref{tab:performance-metrics-low-high} show that the gGAN and the Security Analyzing component are lightweight compared to the enormous latency introduced by the LLM component. Also, longer prompts impose longer latency.

\begin{table}[t]
    \centering
    \tiny
    \begin{tabular}{|l|c|c|}
        \hline
        & \textbf{CPU (Percentage)} & \textbf{Memory (MB)} \\
        \hline
        PromSec & 0.06 & 2,170.75 \\
        \hline
        No PromSec & 0.05 & 2,168.84 \\
        \hline
    \end{tabular}
    \caption{Resource usage comparison} \vspace{-20pt}
    \label{tab:resource-usage}
\end{table}

Third, manually inspecting the code functionality reveals that utilizing a standalone gGAN with commercial LLMs, such as GPT3.5, GPT4, and GPT4o, can cause code functionality deviation, that is, users do not get secure code with desirable functionality given the information loss incurred by the iterative process of optimizing prompts with the LLMs. About 98\% of the generated code has partially or fully deviated functionality. Flexibly replacing the LLMs with a SafeCoder-tuned LLM can potentially enhance the trade-off between code functionality and security. However, these models have been trained on relatively small datasets. Therefore, a more rigorous study is needed to investigate this trade-off and \texttt{SGCode} offers a practical tool for this purpose.

\vspace{-5pt}

\section{Conclusion and Future Works}

We developed \texttt{SGCode}, a timely and flexible system that utilizes prompt-optimizing mechanisms to generate secure code. 
\texttt{SGCode} performs efficiently, even on a lightweight AWS virtual machine, with negligible overhead from prompt-optimizing mechanisms.
Our future work will focus on: 
(1) Incorporating a feature that allows users to define utility tests and automatically test their code; (2) Further optimizing our system to reduce latency; and (3) Scaling the model to handle larger data and more complex code. 

\vspace{-5pt}
\section*{Acknowledgement}

This work is supported by the NSF under the grant CNS-1935928.

\vspace{-2.5pt}


\bibliographystyle{ACM-Reference-Format}
\balance
\bibliography{references}


\begin{thebibliography}{9}


\ifx \showCODEN    \undefined \def \showCODEN     #1{\unskip}     \fi
\ifx \showDOI      \undefined \def \showDOI       #1{#1}\fi
\ifx \showISBNx    \undefined \def \showISBNx     #1{\unskip}     \fi
\ifx \showISBNxiii \undefined \def \showISBNxiii  #1{\unskip}     \fi
\ifx \showISSN     \undefined \def \showISSN      #1{\unskip}     \fi
\ifx \showLCCN     \undefined \def \showLCCN      #1{\unskip}     \fi
\ifx \shownote     \undefined \def \shownote      #1{#1}          \fi
\ifx \showarticletitle \undefined \def \showarticletitle #1{#1}   \fi
\ifx \showURL      \undefined \def \showURL       {\relax}        \fi
\providecommand\bibfield[2]{#2}
\providecommand\bibinfo[2]{#2}
\providecommand\natexlab[1]{#1}
\providecommand\showeprint[2][]{arXiv:#2}

\bibitem[Charalambous et~al\mbox{.}(2023)]%
        {charalambous2023new}
\bibfield{author}{\bibinfo{person}{Yiannis Charalambous}, \bibinfo{person}{Norbert Tihanyi}, \bibinfo{person}{Ridhi Jain}, \bibinfo{person}{Youcheng Sun}, \bibinfo{person}{Mohamed~Amine Ferrag}, {and} \bibinfo{person}{Lucas~C Cordeiro}.} \bibinfo{year}{2023}\natexlab{}.
\newblock \showarticletitle{A New Era in Software Security: Towards Self-Healing Software via Large Language Models and Formal Verification}.
\newblock \bibinfo{journal}{\emph{arXiv preprint arXiv:2305.14752}} (\bibinfo{year}{2023}).
\newblock


\bibitem[Developers(2022)]%
        {bandit_ref}
\bibfield{author}{\bibinfo{person}{B. Developers}.} \bibinfo{year}{2022}\natexlab{}.
\newblock \bibinfo{booktitle}{\emph{Welcome to Bandit — Bandit documentation}}.
\newblock
\urldef\tempurl%
\url{https://bandit.readthedocs.io/en/latest/}
\showURL{%
\tempurl}
\newblock
\shownote{[Online; accessed 1. June 2023]}.


\bibitem[Github(2023)]%
        {copilot}
\bibfield{author}{\bibinfo{person}{Github}.} \bibinfo{year}{2023}\natexlab{}.
\newblock \bibinfo{title}{Research: Quantifying GitHub Copilot’s impact on code quality}.
\newblock \bibinfo{howpublished}{\url{https://github.com/features/copilot}}.
\newblock
\newblock
\shownote{[Online; accessed 17-June-2024]}.


\bibitem[He et~al\mbox{.}(2024)]%
        {he2024instructiontuningsecurecode}
\bibfield{author}{\bibinfo{person}{Jingxuan He}, \bibinfo{person}{Mark Vero}, \bibinfo{person}{Gabriela Krasnopolska}, {and} \bibinfo{person}{Martin Vechev}.} \bibinfo{year}{2024}\natexlab{}.
\newblock \showarticletitle{Instruction Tuning for Secure Code Generation}. In \bibinfo{booktitle}{\emph{ICML}}.
\newblock


\bibitem[Inc.(2021)]%
        {GitHubInc2021CodeQL}
\bibfield{author}{\bibinfo{person}{GitHub Inc.}} \bibinfo{year}{2021}\natexlab{}.
\newblock \bibinfo{title}{CodeQL Documentation}.
\newblock
\newblock
\urldef\tempurl%
\url{https://codeql.github.com/docs/}
\showURL{%
\tempurl}
\newblock
\shownote{[Online; accessed 4-Dec-2023]}.


\bibitem[Nazzal et~al\mbox{.}(2024)]%
        {PromSec}
\bibfield{author}{\bibinfo{person}{Mahmoud Nazzal}, \bibinfo{person}{Issa Khalil}, \bibinfo{person}{Abdallah Khreishah}, {and} \bibinfo{person}{NhatHai Phan}.} \bibinfo{year}{2024}\natexlab{}.
\newblock \showarticletitle{\texttt{PromSec}: Prompt Optimization for Secure Generation of Functional Source Code with Large Language Models (LLMs)}. In \bibinfo{booktitle}{\emph{ACM CCS}}.
\newblock


\bibitem[Pearce et~al\mbox{.}(2022)]%
        {pearce2022asleep}
\bibfield{author}{\bibinfo{person}{Hammond Pearce}, \bibinfo{person}{Baleegh Ahmad}, \bibinfo{person}{Benjamin Tan}, \bibinfo{person}{Brendan Dolan-Gavitt}, {and} \bibinfo{person}{Ramesh Karri}.} \bibinfo{year}{2022}\natexlab{}.
\newblock \showarticletitle{Asleep at the keyboard? assessing the security of github copilot’s code contributions}. In \bibinfo{booktitle}{\emph{IEEE S\&P}}. \bibinfo{pages}{754--768}.
\newblock


\bibitem[Ram{\'i}rez(2024)]%
        {Ramirez_FastAPI}
\bibfield{author}{\bibinfo{person}{Sebasti{\'a}n Ram{\'i}rez}.} \bibinfo{year}{2024}\natexlab{}.
\newblock \bibinfo{booktitle}{\emph{{FastAPI}}}.
\newblock
\urldef\tempurl%
\url{https://fastapi.tiangolo.com}
\showURL{%
\tempurl}


\bibitem[Siddiq et~al\mbox{.}(2023)]%
        {siddiq2023lightweight}
\bibfield{author}{\bibinfo{person}{Mohammed~Latif Siddiq}, \bibinfo{person}{Beatrice Casey}, {and} \bibinfo{person}{Joanna Santos}.} \bibinfo{year}{2023}\natexlab{}.
\newblock \showarticletitle{A Lightweight Framework for High-Quality Code Generation}.
\newblock \bibinfo{journal}{\emph{arXiv:2307.08220}} (\bibinfo{year}{2023}).
\newblock


\end{thebibliography}


\end{document}